# The Large Area Detector onboard the eXTP mission

Marco Feroci*[a,b], Mahdi Ahangarianabhari[c], Giovanni Ambrosi[d], Filippo Ambrosino[a], Andrea Argan[a], Marco Barbera[e,f], Joerg Bayer[g], Pierluigi Bellutti[h], Bruna Bertucci[d,i], Giuseppe Bertuccio[c], Giacomo Borghi[h], Enrico Bozzo[l], Franck Cadeaux[m], Riccardo Campana[n], Francesco Ceraudo[o], Tianxiang Chen[p], Daniela Cirrincione[q], Alessandra De Rosa[a], Ettore Del Monte[a,b], Sergio Di Cosimo[a], Sebastian Diebold[g], Yuri Evangelista[a,b], Qingmei Fan[zz], Yannick Favre[m], Francesco Ficorella[h], Fabio Fuschino[n], Olivier Gevin[o], Marco Grassi[r], Bin Hong[zz], Hanqi Mao[s], Vladimir Karas[t], Tom Kennedy[u], Claudio Labanti[n], Olivier Limousin[o], Ugo Lo Cicero[f], Fang-Jun Lu[p], Tao Luo[p], Piero Malcovati[r], Adrian Martindale[v], Aline Meuris[o], Malgorzata Michalska[w], Alfredo Morbidini[a], Fabio Muleri[a,b], Piotr Orleanski[w], Stephane Paltani[l], Teng Pan[zz], Emanuele Perinati[g], Antonino Picciotto[h], Martin Pohl[m], Irina Rashevskaia[x], Andrea Santangelo[g], Stephane Schanne[o], Konrad Skup[w], Jiri Svoboda[t], Chris Tenzer[g], Andrea Vacchi[q,z], Dave Walton[u], Berend Winter[u], Xin Wu[m], Yupeng Xu[p], Ganluigi Zampa[z], Nicola Zampa[z], Silvia Zane[u], Andrzej Zdziarski[y], Long Zhang[zz], Shu Zhang[p], Shuang-Nan Zhang[p], Wenda Zhang[t], Xiaoli Zhang[zz], Yupeng Zhou[zz], Nicola Zorzi[h]

[a]INAF, Istituto di Astrofisica e Planetologia Spaziali, Via Fosso del Cavaliere 100, 00133 Rome, Italy; [b]INFN, Sezione Roma Tor Vergata, Via della Ricerca Scientifica, 00133, Rome, Italy; [c]Politecnico di Milano, Polo di Como, Via Anzani 41, Como, Italy; [d]INFN, Sezione di Perugia, Via Alessandro Pascoli, 23c, 06123, Perugia, Italy; [e]Università di Palermo, Dipartimento di Fisica e Chmica, Via Archirafi 36, 90123, Palermo, Italy; [f]INAF, Osservatorio Astronomico di Palermo, Piazza del Parlamento, 1, 90134, Palermo, Italy; [g]Institut fur Astronomie und Astrophysik, Eberhard Karls Universitat, 72076 Tuebingen, Germany; [h]Fondazione Bruno Kessler, Via Sommarive, Povo, 38123, Trento, Italy; [i]University of Perugia, Dip. Fisica e Geologia, P.zza Università - 06123 - Perugia Italy; [l]Department of Astronomy, University of Geneva, Chemin d'Ecogia 16, 1290, Versoix, Switzerland; [m]Department of Nuclear and Particle Physics, University of Geneva, CH-1211, Switzerland; [n]INAF, Osservatorio di astrofisica e scienza dello spazio di Bologna, Via P. Gobetti 101, Bologna, Italy; [o]CEA Paris-Saclay, DRF/IRFU, 91191 Gif sur Yvette, France [p]Key Laboratory for Particle Astrophysics, Institute of High Energy Physics, CAS, Beijing 100049, China; [q]University of Udine, Via delle Scienze, 206, 33100, Udine, Italy; [r]University of Pavia, Department of Electrical, Computer, and Biomedical Engineering, Via Ferrata 5, 27100 Pavia, Italy [s]North Night Vision Technology Co. Ltd, Nanjing 211106, China; [t]Astronomical Institute, Czech Academy of Sciences, Bocni II 1401, CZ-14100 Prague, Czech Republic; [u]Mullard Space Science Laboratory, UCL, Holmbury St Mary, Dorking, Surrey, RH56NT,UK; [v]Space Research Centre, Department of Physics and Astronomy, University of Leicester, Leicester, LE17RH, UK [w]Space Research Center, Polish Academy of Sciences, Bartycka 18a, 00-716 Warszawa, Poland [x]TIFPA, Istituto Nazionale di Fisica Nucleare, Via Sommarive 14, 38123 Povo, Trento, Italy [y]Nicolaus Copernicus Astronomical Center, Polish Academy of Sciences, Bartycka 18, PL-00-716 Warszawa, Poland; [z]INFN, Sezione di Trieste, Padriciano 99, 34149, Italy [zz]Beijing Institute of Spacecraft System Engineering, CAST, Beijing 10094, China

*marco.feroci@iaps.inaf.it

**ABSTRACT**

The eXTP (enhanced X-ray Timing and Polarimetry) mission is a major project of the Chinese Academy of Sciences (CAS) and China National Space Administration (CNSA) currently performing an extended phase A study and proposed for a launch by 2025 in a low-earth orbit. The eXTP scientific payload envisages a suite of instruments (Spectroscopy Focusing Array, Polarimetry Focusing Array, Large Area Detector and Wide Field Monitor) offering unprecedented simultaneous wide-band X-ray spectral, timing and polarimetry sensitivity. A large European consortium is contributing to the eXTP study and it is expected to provide key hardware elements, including a Large Area Detector (LAD). The LAD instrument for eXTP is based on the design originally proposed for the LOFT mission within the ESA context. The eXTP/LAD envisages a deployed 3.4 $m^2$ effective area in the 2-30 keV energy range, achieved through the technology of the large-area Silicon Drift Detectors - offering a spectral resolution of up to 200 eV FWHM at 6 keV - and of capillary plate collimators - limiting the field of view to about 1 degree. In this paper we provide an overview of the LAD instrument design, including new elements with respect to the earlier LOFT configuration.

**Keywords:** X-ray Astronomy, Timing, Silicon detectors, capillary plates

## 1. INTRODUCTION

The enhanced X-ray Timing and Polarimetry mission (eXTP) derives from the merging of two major mission concepts in China (XTP[1]) and Europe (LOFT[2]). After having carried out independent phase A studies over the years 2011-2014, supported by the respective funding agencies (CAS for eXTP, ESA and the European Member States for LOFT), the two projects have merged in 2015 in the Sino-European mission concept eXTP. Both XTP and LOFT had been designed to address the science theme of understanding the behavior of matter under extreme conditions of gravity, density and magnetism. Thus, eXTP aims at determining the equation of state of ultra-dense matter in the interior of neutron stars, study the dynamics of matter in the vicinity of neutron stars and near the event horizon in black holes - where the General Relativity theory predicts large distorsions of the space-time with respect to a Newtonian approach – and study the effects of the ultra-critical magnetic fields hosted in magnetar sources on the propagation of photons. These science goals will be achieved through a payload enabling - for the first time ever – high-throughput, simultaneous spectral, timing and polarimetry observations of the same target sources. An extensive description of the eXTP science case is provided in a set of four papers discussing Dense matter[3], Accretion in strong field gravity[4], Strong magnetism[5] and Observatory science[6].

The instruments for these observations are the Spectroscopy Focusing Array (SFA, a set of 9 telescopes equipped with Silicon Drift Detectors, reaching a maximum effective area of 0.7 $m^2$ at 1 keV and <180 eV FWHM spectral resolution at 6 keV), the Polarimetry Focusing Array (PFA, a set of 4 telescopes equipped with the polarization-sensitive Gas Pixel Detectors, reaching an effective area 5 times larger than the polarimeter onboard the IXPE mission[7]), the Wide Field Monitor (WFM, a coded mask instrument simultaneously imaging ~1/3 of the sky, with arcmin resolution[8]) and the Large Area Detector (LAD), the subject of this paper. The eXTP instruments are designed to be fully complementary: the SFA provides the soft response (down to 0.5 keV) and the low background obtained by focusing X-rays onto small detectors, the LAD offers the largest effective area ever achieved at the Fe line energy and the hard response up to 30 keV, the PFA enables the polarization sensitivity in the 2-10 keV range, while the large field of view and angular resolution of the WFM will be crucial for triggering the target sources in their most interesting states and discover new transients. A thorough description of the eXTP mission and scientific payload is provided in Zhang et al.[9,10].

The LAD is designed to be the most sensitive spectral-timing instrument for bright Galactic and extragalactic sources. As we will discuss in the next sections, an innovative and highly efficient technology and design allow to deploy in space an effective area as large as 3.4 $m^2$ at 8 keV, adopting a modular configuration. This is 6x larger than the largest X-ray instruments flown so far, PCA onboard RXTE and LAXPC onboard ASTROSAT, as shown in

Figure 1. In contrast to its predecessors, the LAD combines the large effective area with a spectral resolution of solid state detectors, reaching up to 200 eV FWHM at 6 keV, thus opening for the first time the new window of high-throughput spectral-timing for bright sources. In fact, the largest instrument with similar spectral resolution is XMM-Newton, which has an effective area >30x lower than the LAD and is not optimized for bright sources.

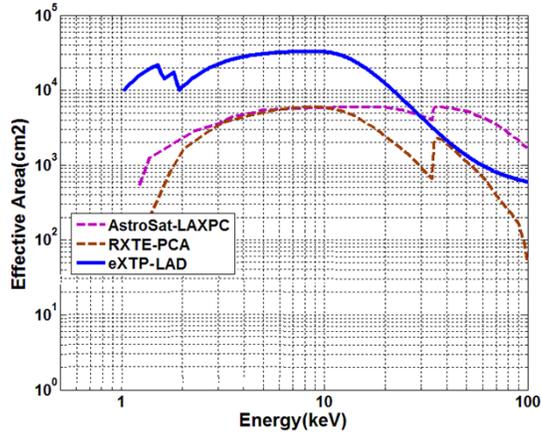

Figure 1. The eXTP LAD effective area, as compared to the past PCA experiment onboard RXTE and the LAXPC experiment onboard the ASTROSAT mission.

The LAD is a large area, collimated instrument (1° field of view, FWHM). This design has the unique advantage of making the dead-time and pile-up effects basically negligible. In fact, each of the ~144,000 channels of the LAD only detects just ~1 count/s even when observing the bright Crab Nebula. As a drawback, the background is higher than in focusing instruments (e.g., the SFA onboard the same payload).

The LAD promises to be a breakthrough in itself for high statistics studies of compact sources, but being also part of the powerful suite of instruments onboard eXTP observing the same sources at the same times, its value will be largely enhanced. In this paper we describe the concept of the LAD instrument, as well as the key technologies that enable its innovative design.

## 2. INSTRUMENT CONCEPT AND DESIGN

The Large Area Detector onboard the eXTP mission is based on the concept earlier developed for the similar LAD instrument onboard the LOFT mission concept[2,11]. The enabling technologies of large-area Silicon Drift Detectors (SDD[12,13]) and capillary plate collimators described in Sect. 3 allow to deploy an unprecedently large collecting area within the boundaries of technical resources of a medium-class mission. The basic conceptual element of the LAD is a sandwich of detector-collimator units. The large-area SDD is the sensitive element, operating in the 2-30 keV energy range. The capillary plate collimator is a Lead-glass tile with thousands of micropores used to limit the field of view to ~1°, in order to reduce source confusion and limit the diffuse X-ray background. The small volume and light weight of both elements (the 120x72x0.45 mm$^3$ detector weighs ~10 grams, while the capillary plate collimator - with the same size of the SDD and a thickness of 5 mm – has a mass of ~50 grams) allow to develop an instrument with a highly efficient and innovative mass per unit surface, about ten times lower than earlier designs based on traditional proportional counters and metal collimators.

The LAD instrument is organized in Modules. Each Module is composed of a set of 16 detectors and 16 collimators, assembled in two independent grids, that are then bolted together. Each Module hosts its own Module Back End Electronics (MBEE), including the control and I/O electronics for the detectors, as well as their power supplies. The Modules are then organized in large Panels – 20 Modules per Panel - which are deployed from the satellite structure. Each Panel includes two Panel Back End Electronics (PBEE), one serving 10 Modules, in charge of interfacing the Modules to the Instrument Control Unit (ICU). In Figure 2 the hierarchical concept of the LAD instrument is shown, while Figure 3 shows its architecture. Being composed of independent Modules, each of them with its own electronics, the LAD modular design makes it intrinsically highly redundant, with the potential loss of one Module not affecting the rest of the instrument. Overall, the LAD instrument onboard eXTP is composed of 2 Panels, including 40 Modules, including 640 detector-collimator units, reaching up the effective area shown in

Figure 1 as a function of energy.

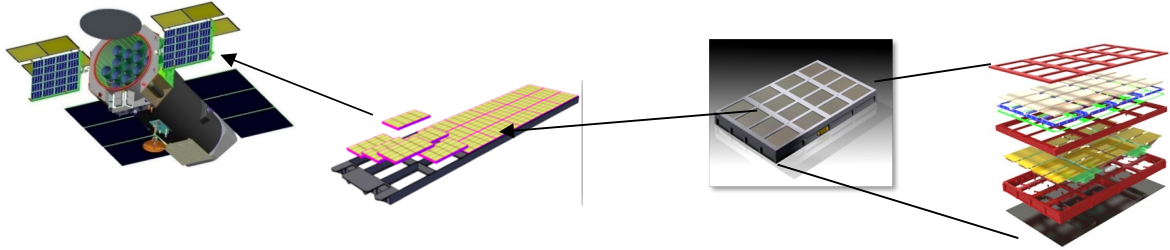

Figure 2. The hierarchical structure of the LAD instrument. From the left: the eXTP satellite with the deployed LAD Panels (blue, with their Sun shield), the conceptual structure of the LAD Panel (the eXTP configuration has 4x5 Modules in each of the 2 Panels), the LAD Module (size is about 330 x 540 mm$^2$) and its exploded view (top to bottom: collimator frame, collimator plates, clamps, SDDs+FEEs, detector frame, radiator). The satellite design shown here is the current baseline from the CAST company.

The LAD is a collimated instrument, with a field of view of 1 degree FWHM. As such, the different Modules are required to be co-aligned. Any misalignment would effectively result in a smaller effective area and a larger field of view. The alignment requirement and strategy affects and drives several elements in the design. The field of view is limited by the individual pores of the capillary plates, which are then required to be co-aligned. The mounting of the 16 capillary plates within the Module collimator frame does not foresee individual fine-adjustments. This implies a requirement to the perpendicularity of the pore directions to the surface of the capillary plate, which is then mechanically interfaced ("plug-and play") to its hole in the collimator frame. The 16 capillary plates inside a Module are also required to be co-aligned one another. Similarly, the 40 Modules within a Panel have to guarantee a proper co-alignment, withstanding the launch loads and thermo-elastic distortions. Finally, the deployment system of the two Panels and their structure are required to guarantee the co-alignment and stability of the two panels.

The alignment of the LAD elements is important to preserve the effective area within requirements. However, this element has to be traded-off against the response stability of the instrument. In fact, the ideal collimator response is a nearly triangular function: any instabilities of the attitude control system of the spacecraft will effectively result in a linear variation of the effective area of the instrument, as the effective instantaneous pointing will lay at different locations within the field of view (that is, at different position of the triangular response of the collimator). The resulting effect would be a spurious modulation in the source counting rate, due to a modulation of the instrument effective area rather than to the intrinsic source variability. Preventing this effect, by reducing it below the sensitivity limit of the instrument, is achieved by combining the alignment requirements of the instrument with a response stability requirement of the attitude and orbit control system of the satellite. The latter is implemented as a function of frequency, as the sensitivity of the instrument, as well as the expected variability of target sources, is different as a function of frequency.

The overall alignment budget of the LAD is 4.6 arcminutes (99.7% confidence), which is then broken down and assigned to the individual components mentioned above, while the response stability requirement is specified in Table 1 for different frequency ranges.

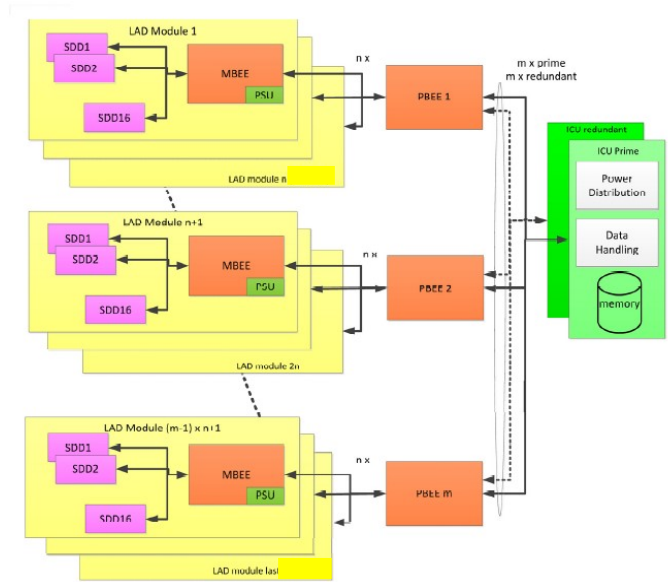

Figure 3. The architecture of the LAD instrument. Each Module is composed of 16 detector units and one MBEE. One PBEE interfaces 10 Modules and all PBEEs are interfaced by a single ICU, which is cold-redundant.

Table 1. The main specifications of the LAD instrument

| Parameter | Specification |
|---|---|
| Energy range | 2-30 keV (extended: 30-80 keV for out-FoV) |
| Effective area | 3.4 m$^2$ @8 keV, 0.9 m$^2$ @2 keV, 0.3 m$^2$ @30 keV |
| Energy resolution | <240 eV FWHM @6 keV (all events) |
| Field of View | 1° (FWHM) |
| Field of Regard | >50% |
| Absolute Time Accuracy | <10 µs |
| Deadtime | <1% @1 Crab |
| Background | <10 mCrab |
| Background knowledge | <0.5% at 5-10 keV |
| Response stability | <0.01 Hz: <2% per decade<br>0.01-1 Hz: <0.2% per decade<br>1-1200 Hz: <0.02% per octave<br>>2000 Hz: lower is better<br>10-2000 Hz: <0.0002% nearly periodic |
| Maximum flux | >500 mCrab (sustained)<br>>15 Crab (continuous, 300 minutes) |

It is worth noticing that, in contrast to the earlier LOFT design, the LAD instrument onboard eXTP has to operate simultaneously with the SFA and PFA instruments. As a consequence, the overall alignment of the LAD has to point its maximum effective area in a direction co-aligned with the boresight of these other two instruments. This imposes a further alignment requirement between the LAD and the nominal boresight of the spacecraft and will require a specific calibration on ground and in orbit.

**2.1 Detector Assembly**

The LAD Detector Assembly is composed of the SDD detector and its front-end electronics (FEE) board. The PCB hosting the FEE components also acts as mechanical support to the SDD and as interface to the Module structure. Figure 4 shows the SDD (before cutting it out of its production wafer at the manufacturer site) and the Detector Assembly design. The H-shaped support structure providing the mechanical interface to the Detector Module Frame is visible on top, with its 4 interface screws. The SDD is glued to the PCB on its back-side, where the charge collecting anodes are located. The size of the PCB is intentionally narrower than the SDD thus allowing a direct wire-bonding of the SDD anodes to the inputs of the read-out ASICs.

The read-out of the 76-cm$^2$ monolithic LAD detector is based on the IDeF-X HD ASIC design[14]. Eight 32-channel IDeF-X HD ASICs interface the 224 anodes of the SDD (112 on each SDD side). Each IDeF-X HD channel provides the full analogue chain for the read-out of the charge signals from the SDD. The analogue-to-digital conversion is then provided by one 16-channel OWB-1 ASIC[15]. In addition to the ASICs, the FEE board includes all the required filters and biases for both the ASICs and the detector. The board is designed on a rigid-flex PCB to incorporate the interconnection cables to the MBEE, for signals and power supplies (see Figure 4). This solution is adopted both for saving space on the PCB and to ease the assembly procedure of the Detector Assembly into the Module. A mechanical breadboard was prototyped at DPNC (Univ. of Geneva) during the LOFT study to test the assembly procedures and interconnections.

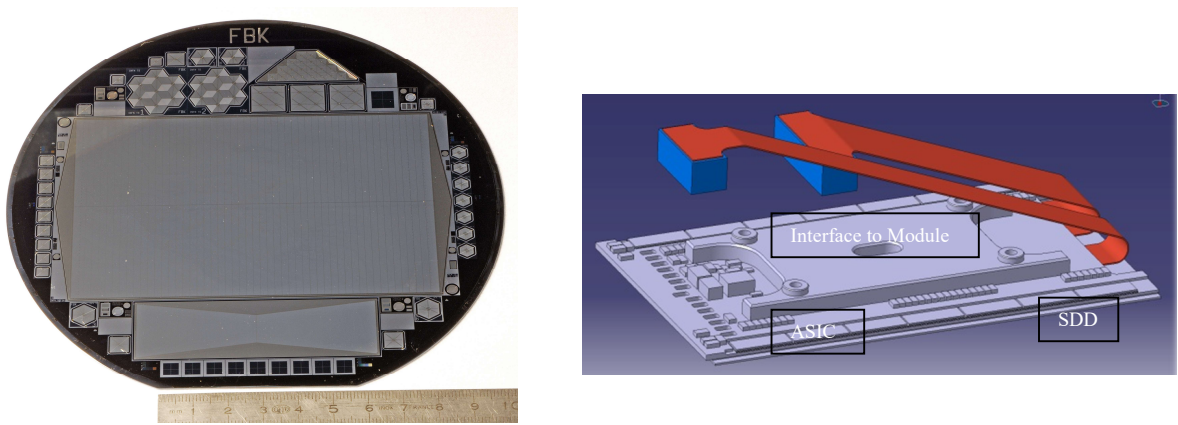

Figure 4. *Left*: the large-area Silicon Drift Detector (still on its 6" production wafer, together with other detectors and test structures). The triangular voltage dividers on the two sides of the SDD are clearly visible. *Right*: CAD drawing of the Detector Assembly, including the SDD and its front-end electronics.

**2.2 Detector Module**

The Detector Module (DM) is the basic assembly unit of the Large Area Detector. It is in charge of supporting and interfacing 16 capillary plates collimators, 16 detectors with their front-end electronics and the Module Back End Electronics, providing power supplies, configuration biases and I/O interfaces. The DM is meant to be the minimum self-consistent, modular element of the LAD. The full instrument includes 40 DM, organized in two deployable Panels.

The structure and components of the DM are shown in Figure 5. It is composed of two main elements: a Detector Frame and a Collimator Frame. The Detector Frame is the mechanical support structure for the 16 Detector Assemblies and the MBEE. The Detector Assembly is the sensitive element of the LAD but being a collimated (as opposite to imaging) instrument, it has no special requirements for the alignment. Instead, it has significant design requirements related to the mass budget, power budget, volume and AIVT procedures. The LAD is composed by a large number (40) of DMs. This requires efficient "mass production" and AIVT. The DM was therefore designed up-front to be as much as possible simple and "plug-and-play", keeping in mind that any complexity in the assembly or test procedures would translate into a long AIVT phase and a significant impact on the overall project schedule. Similarly, considering the large number of DMs, the volume and the mass budget, as well as the power budget, of each element multiplies by 40. So, any small increase in the technical budget of the DM immediately implies a large increase in the technical budget of the LAD instrument, especially for the mass and volume, for which larger values would also require larger and heavier Panels.

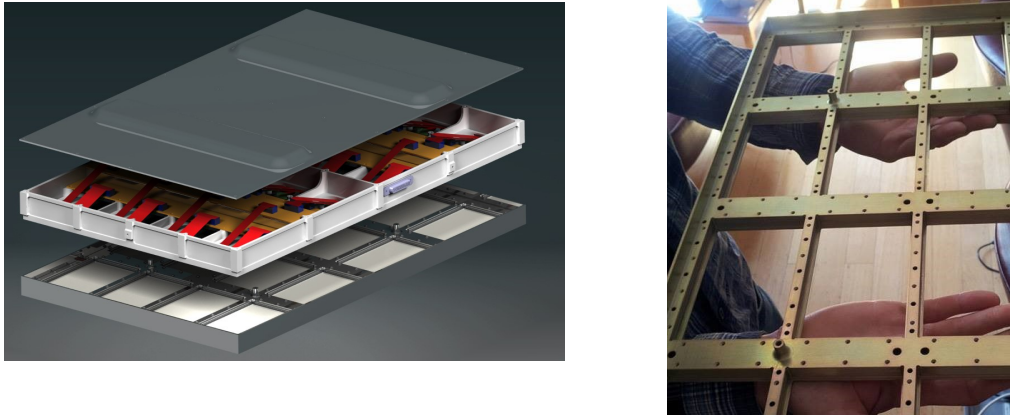

Figure 5. *Left:* CAD drawing of the Detector Module showing an exploded view. The bottom structure is the Collimator Frame, hosting and supporting the 16 capillary plates and optical-thermal filters. The middle structure is the Detector Frame, interfacing the 16 Detector Assemblies and hosting the Module Back End electronics. The top cover is the back-side radiator, also supporting the Lead shield. Right: a prototype of the Collimator Frame that was used to test the assembly procedure of the capillary plates and to verify them through vibration tests.

The power budget is also an important design element for the LAD, under two respects. As mentioned above, the power budget of each DM has to be kept under control, as it multiplies by 40, with immediate impacts on the overall payload power budget. On the other hand, as we will discuss in Sect. 3.1, the SDD require a low operating temperature, to compensate for the predicted radiation damage in orbit. Considering the large size of the LAD, cooling can only be passively achieved by a proper thermal design. The power required locally in the DM therefore contributes to the LAD thermal design with locally generated heat, that has to be removed. A low DM power budget is therefore important to make the LAD thermal design more efficient. In this context, the backside cover of the DM is designed as a thermal radiator. In the current baseline design, the power budget for the DM is 8.3 W CBE, including DC-DC conversion (in-)efficiencies.

The Collimator Frame is the structural element for the 16 capillary plates of a DM. The collimators are the optical elements of the LAD, therefore the mechanical interface provided by the Collimator Frame is also in charge of providing their co-alignment, as well as the required rigidity and mechanical stability over the expected temperature variations. The 16 capillary plates are hosted in recesses in the Collimator Frame, designed for a plug-and-play assembly, with the alignment being guaranteed by the flatness of the mechanical interface. The integration procedure foresees the 5 mm thick capillary plates to be integrated first and then the 1 µm thick optical-thermal filters piled on top of them. A narrow portal frame pressure plate is placed on top of each plate and then a single stainless steel part with integral springs for each plate so that the spring fingers straddle the collimator frame bars and touch down on the backs of the pressure plates. A simple clamp plate is then placed on top of the spring plate and evenly screwed down to compress the springs while the front collimator frame face is held against a surface table. The spring guarantees sufficient preload, adequately spread on the collimator tile to prevent it from moving out of plane during vibration testing. The lateral movement of the tiles is limited by the springs as well using 4 clips that are part of the stainless steel spring, pushing against each edge and centering it within each aperture. Hence the lateral movement is constrained by the spring itself and does not rely on the out of plane preload alone. This system makes for faster assembly while permitting sufficient lateral movement to handle CTE differences. This clamping system was preliminarily tested by producing a prototype of the Collimator Frame, integrating it with dummy collimator tiles and (successfully) vibration testing it (see Figure 5, right). In order to save mass, this frame features hollow tubular box section members to give maximum stiffness for minimum weight and is fabricated from machined channel section halves that are then dip brazed together and finish machined.

Finally, the DM also hosts a 300 μm thick Lead shield, glued to the backside radiator. This is needed to shield the detectors from the diffuse Cosmic X-ray Background hitting from the back, as the deployed panels are basically a support grid and have no heavy structure that could shield X-ray radiation.

Based on the current baseline design of the DM, the mass budget is 6.1 kg (CBE) per Module.

### 2.3 Detector Panel

The 40 LAD Modules are supported and deployed in space through Detector Panels (DP). These are grid-like structures that are required to guarantee support and co-alignment of the Detector Modules among them and with the satellite boresight. Considering their large size and mass, the Detector Panels are an integral part of the spacecraft design and for this reason they are in charge to the prime contractor of the spacecraft, that will design them coherently with the overall design solutions. Currently, two Chinese companies are studying the eXTP spacecraft and systems in parallel, CAST (Beijing) and Micro-Sat (Shanghai). The final implementation of the Detector Panels will thus depend on which one of the two studies will be adopted. However, in both cases the DP design will have to guarantee adequate stiffness at minimum weight. The preliminary design envisages the use of Carbon Fiber Reinforced Plastic as structural material.

The DP are also part of the overall thermal design of the LAD. This will have an impact on both the selection of material, as well as on their design. In this respect, a trade-off study will be performed on the efficiency of having a thermal radiator on the backside of each Module or a single common radiator on the backside of the DP. In any case, the required low operating temperature of the LAD detectors suggest avoiding direct illumination by the Sun. To achieve this, light and deployable sun shields are currently envisaged for the LAD Panels (see Figure 2).

The DP host the Panel Back-End Electronics (PBEE) boxes of the LAD. These are 4 interface electronic boards between the MBEE and the single, central Instrument Control Unit of the LAD. In principle, the location of the PBEE on the spacecraft is unconstrained. The routing of the digital signals they require is not particularly sensitive to their position and/or cable length. However, each PBEE interfaces 10 MBEEs (Modules) with a consequent large bunch of cables. In this respect, placing the PBEE on the rear of the mechanical structure of the Panel has the major advantage of not having this bunch of wires crossing the hinge of the Panel. On the other hand, the presence of the PBEE boxes is in itself a break in the Panel thermal symmetry and this has an impact on the overall thermal design. A trade-off study for this issue is thus being carried out by the industrial primes.

### 2.4 The LAD digital electronics: MBEE, PBEE and ICU

Following the modular philosophy of the LAD design, the digital electronics is split in modular units as well. Each of the 40 Detector Modules hosts its MBEE, whereas each of the 2 panels hosts 2 PBEE. The 4 LAD PBEE interface the single ICU, which is cold-redundant. The MBEE, the PBEE and the ICU have been prototyped during the LOFT study to test their functionalities, software codes and interfaces. The prototypes are shown in Figure 6.

*MBEE – Module Back End Electronics*

When the charge collected at one anode exceeds a programmable threshold in one of the ASIC channels in the Front-End Electronics (FEE), a trigger signal is forwarded to the MBEE where a time tag (based on the 1 MHz clock count provided by the ICU) is instantaneously generated. The trigger is also propagated directly from one ASIC to all the ASICs on the respective detector half and they all freeze their current signal. The MBEE requests the trigger map from all the ASICs on this detector half and validates if only one or two adjacent anodes triggered. If the event passes this selection criteria, the "A/D conversion" command is sent from the MBEE to the ASIC and the conversion is carried out (inside the OWB-1 ASIC), providing an 11-bit output per anode.

Following the A/D conversion, the MBEE processing pipeline will be activated. The saved time tag will be added to the event package at the end of the processing pipeline. The main processing functions of the MBEEs are:

- time tagging
- trigger validation and filtering

- pedestal subtraction
- common noise subtraction
- energy reconstruction
- event threshold application
- housekeeping data collection.

Each MBEE consists of two PCBs and is designed to process events within a pipeline structure that handles the events from one of the two sides of a detector. This pipeline is initiated 32 times within the MBEE FPGAs to allow processing of data from all 16 detectors simultaneously. The pipeline is designed such that the processing time within each step is the same and shorter than the A/D conversion time of a following event. In this way, the data processing in the MBEE does not inflict any additional dead-time and the pipeline is always ready for the next event.

The first step within each pipeline is the pedestal subtraction where a set of pedestal values (one for each anode) stored within the MBEE are subtracted from the measured signal values. Then, the common noise (CN) is calculated and also subtracted. The CN is a noise component common to all the channels connected to the same ASIC and basically an undesired baseline shift. The CN will be calculated independently for each ASIC considering only the channels not hit by the charge cloud. The next step of the pipeline concerns the on-board energy reconstruction. The procedure will be composed of the following steps: 1) Gain Correction: the reconstructed energy of the event is the sum of the n event channels, each one multiplied by an individual gain factor that is kept in a lookup table which in the same fashion as the pedestal table can be recalculated or uploaded from ground. 2) Temperature Correction: the gain calibration will be performed at a fixed temperature and a gain variability factor per degree is assumed. A linear correction is applied to the energy value of the event. 3) Upper Threshold Rejection: if the reconstructed energy exceeds an upper threshold, the event is rejected. 4) Energy Scaling: the last step of the on-board energy reconstruction generates the final 9 bit energy word according to a non-linear function as follows: 2 keV – 30 keV: 60 eV per digit; 30 keV – 80 keV: 2 keV per digit. The TM/TC interface between MBEE and PBEE is using the SpaceWire hardware standard but does implement a custom SPI-like data transfer running at 10 MHz.

*PBEE – Panel Back End Electronics*

The Panel Back-End Electronics (PBEE) handles all events from the individual modules of one of the two detector panels. It is the heart of the data acquisition and signal processing, located between the individual modules and the Data Handling Unit (DHU, part of the ICU) on the satellite bus. The main tasks of the PBEE are:

- Interfacing the MBEEs
- Collecting and buffering the event packets
- Differential time assignment
- Reformatting the data to binned data depending on the observation mode
- Transferring the data to the DHU
- Collection of housekeeping (HK) data and creation of HK packets.

The interface between PBEE and DHU is a fully compliant SpaceWire interface, runnnig at 100 MHz.

*ICU – Instrument Control Unit*

The Instrument Control Unit is the main controlling element of the LAD instrument. It provides the interface to the spacecraft OBDH and it controls all LAD instrument sub-systems via SpaceWire. The ICU includes three components: the Data Handling Unit (DHU), the mass memory and the Power Distribution Unit (PDU). The tasks performed at the level of the ICU involve telecommand execution and distribution, access to mass memory, time distribution, data processing and compression, HK collection, instrument health monitoring and calibration tasks.

The ICU will manage the telemetry generated by the LAD instrument. Each LAD event will require 24 bits: event ID (3 bits), differential time (12 bits) and energy (9 bits). Transmitting all events to the ground, the instantaneous telemetry rate generated by the LAD will of course depend on the specific intensity of the target source. However, the average telemetry allocation for the LAD has been dimensioned by considering the minimum requirement of a sustained observation of 500 mCrab sources. In this case, the telemetry rate generated by the LAD is 654 kbps, of which 610 kbps are due to event data and the remaining part accounts for scientific ratemeters and housekeeping data.

As a single point of failure, the ICU is in cold redundancy. At its heart is the Data Handling Unit where the scientific data is compiled, processed and compressed. A quad-core GR-740 Leon-4 processor (not under ITAR restrictions) was chosen for the task to run the onboard software due to its additional flexibility when compared to a hardware only architecture, i.e. a state-machine. The DHU collects the data from the PBEEs via a SpaceWire interface PCB, performs the selection and formatting tasks depending on the selected observation mode and commits the data to the mass memory in a compressed format. Besides the routing and execution of telecommands, the ICU can run macro commands for several tasks, including on-board calibration and the individual sequenced switching of the MBEEs in the engineering modes.

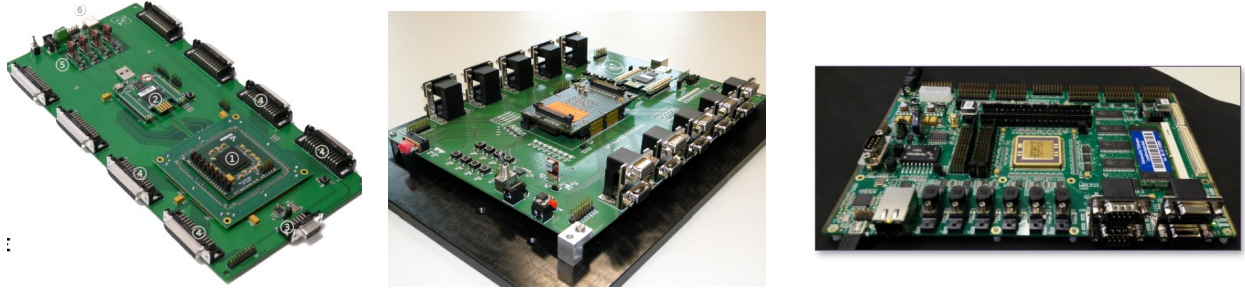

Figure 6. Prototypes of the digital electronics of the LAD developed by the University of Tuebingen: MBEE (left), PBEE (center), ICU (right).

## 3. KEY TECHNOLOGIES

The design of the LAD is based on a set of key technologies, that were and are being developed within the eXTP consortium: the large-area Silicon Drift Detector, the capillary plate collimator, the read-out ASICs and the optical-thermal filter. In this section we briefly review the status of these LAD components.

### 3.1 Large area Silicon Drift Detector

The large-area SDDs were originally developed for the ALICE Inner Tracking System CERN LHC by INFN Trieste, in co-operation with Canberra Inc.[12,16]. In ALICE, 260 SDDs, for a total area of 1.37 $m^2$ are successfully operating since 2008. The key properties of the Si drift detectors[17] are their capability to read-out a large photon collecting area with a small set of low-capacitance (thus low-noise) anodes and their very small volume (450 µm thickness) and low weight (~1 kg $m^{-2}$). The working principle is as follows: the cloud of electrons generated by the interaction of an X-ray photon is drifted towards the read-out anodes, driven by a constant electric field sustained by a progressively decreasing negative voltage applied to a series of cathodes, down to the anodes at ~0 V. The diffusion in Si causes the electron cloud to expand by a factor depending on the square root of the drift time. The charge distribution over the collecting anodes then depends on the absorption point in the detector.

The LAD detector design is an optimization of the ALICE detector, developed in a collaboration between INFN-Ts, Fondazione Bruno Kessler (FBK) and INAF. 6-inch, 450 µm thick wafers are used to produce 76 $cm^2$ monolithic SDDs. The anode pitch is increased to 970 µm to reduce power consumption. The Si tile is electrically divided in two halves, with 2 series of 112 read-out anodes at two edges and the highest voltage along its symmetry axis. The drift length is 35 mm. A drift field of 370 V/cm (1300 V maximum voltage), gives a drift velocity of ~7 mm/µs and a maximum drift time of ~5 µs, the highest detector contribution to the uncertainty in the photon arrival time measurement, a factor of 2 smaller than the LAD scientific requirement. Indeed, with the above electric field, the charge will often distribute over 2 anodes, depending on the photon absorption point: approximately 45% of the events will be single-anode and 55% will be double-anode detections. Detector segmentation makes dead time and pile-up negligible, a crucial property for a timing experiment.

Several versions of the LAD SDD have been developed during the LOFT study[13] and afterwards by R&D activity supported by INFN and the Italian Space Agency. Figure 7 shows the full-size prototype of the large area SDD and its spectral resolution performance, achieved in the lab with a breadboard equipped with the VEGA ASIC[18]. Qualification campaigns have also been performed during the LOFT study: proton irradiation tests at the Paul Scherrer Institute for radiation damage due to Non-Ionizing Energy Losses effects[19] and for Charge Collection Efficiency[20] (CCE); irradiation with low energy protons (200 keV and 800 keV) at the accelerator at EKUT to study the potential damage to shallow implants[21]; irradiation with debris at the Van der Graf accelerator at MPIK in Heidelberg to study the damage due to micro-meteoroid impacts[22]. Radiation damage due to dose effects had been tested already before LOFT[23]. The SDDs successfully passed all tests, demonstrating its qualification for the radiation environment in low-Earth orbits. The SDD were also thermally tested to study operation and performance as a function of temperature and verify the recovery action of a low operating temperature with respect to radiation damage (the leakage current halves every 7°C), as reported in several studies in literature for Silicon detectors.

The radiation tests confirmed the susceptibility of the SDD detectors to NIEL damage induced by protons in orbit. The effect is a displacement damage which causes an increase in the bulk leakage current. The latter is one of the main components of the spectral performance and can only be compensated by lowering the operating temperature (or by annealing procedures). A study of the radiation environment for orbits at different altitudes and inclination has shown that the best orbit for the LAD is equatorial, at an altitude of 550 km. This reduces the radiation damage induced by protons trapped in the Van Allen belt at its minimum. Considering the radiation damage received over 5 years in orbit ("end of life" condition), the SDD will be able to satisfy their spectral resolution requirements if operated at a temperature of -11°C or lower. The orbit and the operating temperature were thus provided as system-level requirements to the eXTP spacecraft.

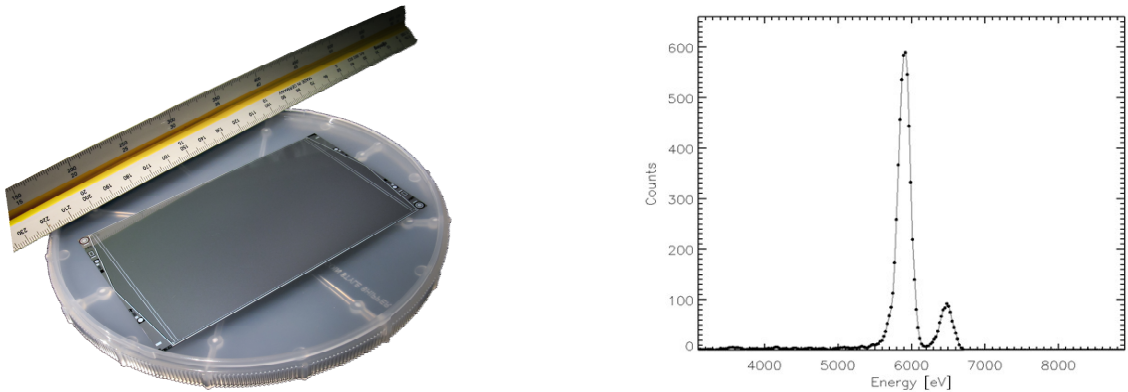

Figure 7 *Left:* The full-scale prototype of the large-area SDD, developed in Italy by a collaboration among INFN, FBK and INAF. Right: an energy spectrum (Fe55 radioactive source, lines at 5.89 and 6.49 keV) taken at leakage current value representative of the end-of-life conditions. The FWHM energy resolution for single-anode events is about 200 eV.

## 3.2 ASICs

Each of the 16 SDDs in a Module is equipped with its own FEE board. For the high-density read-out of the detector, dedicated ASICs with high performance and low power (17 e- rms noise with 650 µW/channel) are needed. The read-out is performed by eight 32-channel IDeF-X HD ASICs, with A/D conversion carried out by one 16-channel OWB-1 ASIC for every detector. Figure 8 shows pictures of the current versions of the IDeF-X and OWB-1 ASICs.

The design of the ASIC is based on previous developments by CEA-IRFU for the readout of CdTe detectors for hard X-ray spectroscopy, i.e. the IDeF-X ASICs (Imaging Detectors Front-end for X-rays[24,14]). IDeF-X HD is the last version of the generation that was successful space qualified and implemented in detector systems (Solar Orbiter STIX experiment). It is produced by AMS in CMOS 0.35 µm technology. It includes 32 analogue channels and a common part for slow control and readout communication with a specific controller. The analogue channel is a charge sensitive preamplifier (CSA) with a continuous reset system followed by a variable gain stage to select the input dynamic range, a pole zero cancellation implemented to avoid long-duration undershoots at the output and to perform a first integration and a second order low-pass filter (RC²) with variable shaping time. The output of each analogue channel feeds a discriminator that compares the amplitude with an in-pixel reference low level threshold to detect events, and a stretcher made of a peak detector and a storage capacitor to sample and hold the amplitude of the signal which is proportional to the integrated charge and hence to the incident energy. The power consumption is very low (850 µW/channel measured at 3.3V).

The CEA team recently designed a new ASIC called IDeF-X HDBD without protection diodes at the inputs to strongly decrease the equivalent noise charge as compared to IDeF-X HD. This incremental improvement is known to be not yet enough to satisfy the noise performance requirement of the LAD (25 electrons in simulation for a dark current of 5 pA, which is the expected anode current of both detectors at the end of life), however it is already a major (~2x) improvement from the original IDeF-X design and its tests will provide key information towards further improvements. The trigger logic has also to be revised to allow synchronous data sampling in the channels neighbouring the main hit channel. Finally, the current 4-side pad distribution will need to be re-designed in a 2-side configuration, to allow for side-by-side assembly in front of the SDD anodes.

The OBW-1 ASIC, also developed at IRFU/CEA Saclay, is devoted to the analogue-to-digital conversion of IDeF-X HD analogue output signals. The OWB-1 is a 14-bit parallel Wilkinson ADC with 16 differential input channels, low conversion time (<2.8 µs) and low power consumption (<2 mW/differential input channel). It is built by AMS (CMOS 0.35 µm technology) and is radiation hardened by design. In 2014 a first version of this chip has been produced and passed successfully all functional and performance tests[15]. In addition, OWB-1 passed successfully the critical latch-up tests.

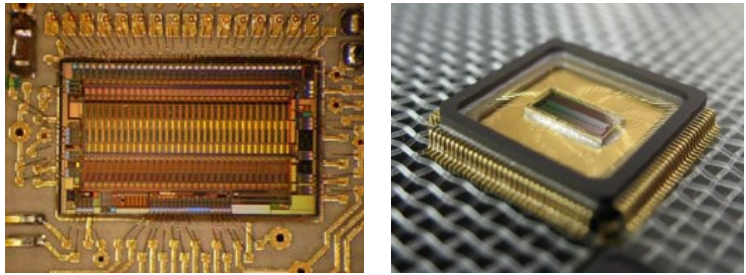

Figure 8. The LAD ASICs, developed by in France by CEA. Left: the IDeF-X HDBD on its test board. Right: the OWB-1 ASIC bonded in its test package.

### 3.3 Capillary plate collimators

To get full advantage of the low weight and small volume of the large-area SDD, the collimation of the field of view is implemented through the technology of capillary plates. This is the same technology as the substrate of the micro-channel plates: a thin sheet of Lead-glass with thousands of micro-pores. Actually, the fabrication of these devices is done by repeated drawings and fusions of "micro-fibers" which eventually form a cylindrical ingot with the required section, that is then sliced to obtain the thin plates. This technology is already long space-proven, not only through microchannel plates detectors (e.g., EXOSAT, Chandra, …) but also as X-ray collimators (EXOSAT Medium Energy and shortly the MIXS-C experiment onboard BepiColombo).

In the LAD configuration, the capillary plate has the following requirements:

- Lead content: >37%
- Size: 111 mm x 72.5 mm
- Thickness: 5 mm
- Aspect ratio: 60:1
- Pore size: 83 µm diameter, round
- Open Area Ratio >70%
- Low contamination by radioactive elements (e.g., $^{40}$K)
- Pore to pore alignment: 1 arcmin
- Pore to surface orthogonality: 1 arcmin
- Surface roughness of inner pore walls: > 13 nm rms.

For eXTP the technology of North Night Vision Technology (NNVT, Nanjing, China) is currently under investigation. Several prototypes have been developed and tested over the past few years[25] and recently a nearly full-size prototype has been fabricated, with a 75% open area ratio, a Potassium content <0.05% and a surface roughness >50 nm rms. In the very near future the full-size will be achieved and it will be used for assembly test with prototypes of the Collimator frame. In Figure 9 we show a photo of the latest prototype, with a zoom-in of the pores. During the LOFT study, capillary plates technology developed by Photonis (France) and Hamamatsu Photonics (Japan) were also investigated and tested. Both of them proved to be suitable for the LAD application. In particular, Photonis produced a full-spec prototype under an ESA contract.

The collimating properties of the NNVT capillary plate has been tested at the X-ray facility of INAF-IAPS Rome. In Figure 10 we show the rocking curve (count rate as a function of the rocking angle) over the entire field of view, at 6.4 keV and 17.5 keV. As shown by the data, at 17.5 keV the angular response is nicely triangular, implying a good stopping power of the Lead-glass, as well as a good pore-to-pore alignment. The zero-response field of view is ±1°, as designed. At 6.4 keV the data show traces of X-ray reflection of the inner pore walls (the response is a bit wider than triangular and rounded at the top). As the X-ray reflection increases at decreasing energy, the result at 6.4 keV is consistent with that at 17.5 keV. However, it also indicates that a significantly larger reflection is to be expected at lower energies. This is currently being tested and will be used as a feedback to the manufacturer for improving the inner pore surface.

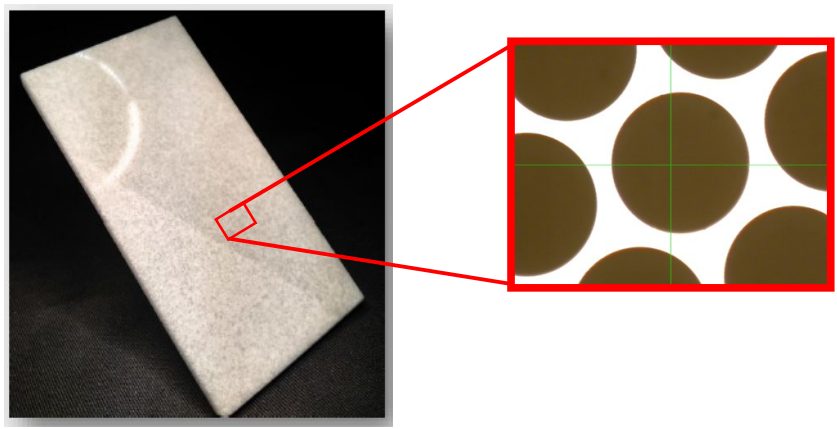

Figure 9. The capillary plate prototype fabricated by NNVT for the LAD

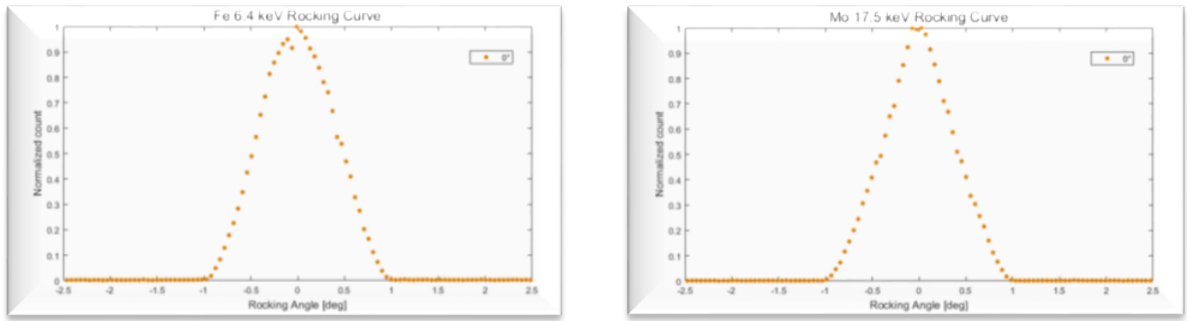

Figure 10. The angular response (rocking curve) of the NNVT capillary plate as tested at the X-ray facility of INAF-IAPS in Rome, at an energy of 6.4 keV (left) and 17.5 keV (right).

### 3.4 Optical-thermal filter

The optical-thermal filter is a key element for the LAD instrument, playing the double role of guaranteeing the light-tightness of the detectors and participate in the thermal design of the Module. The requirement for light tightness is $10^{-6}$ transmission at visible wavelengths, while guaranteeing >90% transparency at 2 keV. The filter for the LAD is being developed at IHEP (Beijing), on the heritage of the similar filter currently flying onboard the HXMT mission[26]. The baseline design for the LAD is a 1 µm support substrate in Polymide, with a 120 nm Aluminum layer deposited on one side.

The development of this filter is quite advanced. Full-size prototypes have been fabricated and will undergo both physical (e.g., transmission as a function of wavelength, from IR to X-rays) and environmental (e.g., thermal, acoustic, …) tests in the near future. The full-size prototype and preliminary transmission tests from IR to UV are shown in Figure 11.

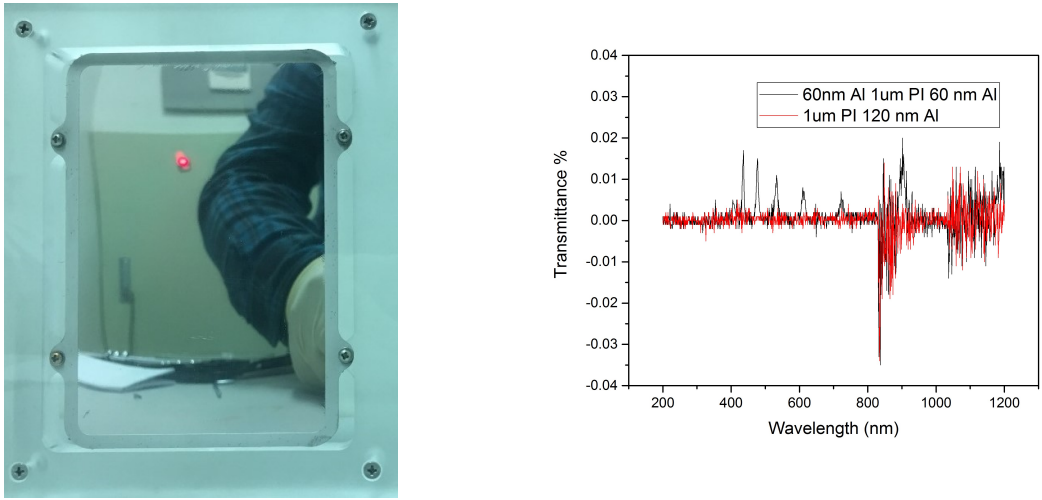

Figure 11. Left: The full-scale prototype of the LAD optical-thermal filter manufactured at IHEP on the heritage of the similar filter onboard HXMT. The supporting material is 1 µm thick Polymide, while the coating is 120 nm Aluminum. Right: test of the transmission from UV to IR for two different filters (the Aluminum coating on both sides or on a single side).

## 4. CONCLUSIONS

The Large Area Detector will be one of the European contributions to the eXTP mission, under the responsibility of the European eXTP Consortium. In particular, the LAD design and development is currently coordinated by the Italian Space Agency (ASI), with significant contributions from Germany, France, Switzerland, Poland, Czech Republic and the UK in Europe and from China. The current LAD design builds on the LAD design derived from the LOFT-M3 study.

The eXTP project is currently undergoing an extended Phase A study in the context of the Chinese Academy of Sciences. The study recently received funding support up to the completion of Phase B and the beginning of Phase C in 2020. The current schedule foresees a launch around 2025. ASI recently formally joined the mission development and will act as coordinating entity of the European eXTP Consortium. Discussions are ongoing concerning a potential involvement of the European Space Agency.

## ACKNOWLEDGEMENTS

The SDD development results described in this paper were obtained within the INFN project ReDSoX2 (Research Detectors for Soft X-rays), also supported under ASI agreement 2016-18-H.0, INAF grant TECNO-INAF-2014, FBK-

INFN agreement 2015-03-06. The Italian study team also acknowledges support by the Italian Space Agency. The Chinese team acknowledges support by the Chinese Academy of Sciences through the Strategic Priority Research Program, Grant No. XDA15020100. The German team acknowledges support from the Deutsche Zentrum fur Luft- und Raumfahrt, the German Aerospce Center (DLR). The Polish Team acknowledges the support of Science Centre grant 2013/10/M/ST9/00729.